\documentclass[seceq]{ptptex}
 
\notypesetlogo                       
\preprintnumber[3cm]{arXiv:physics/0307008}

\title{
Thermodynamics in Density-Functional Theory\\
 and Force Theorems
}

\author{
Junzo \textsc{Chihara}$^{1,}$\footnote{E-mail: chihara@ndc.tokai.jaeri.go.jp} and Mitsuru \textsc{Yamagiwa}$^{2}$%
}

\inst{
$^1$Research Organization for Information Science and Technology,\\
Tokai, Ibaraki 319-1106, Japan\\
$^2$Advanced Photon Research Center, JAERI,
Kizu, Kyoto 619-0215, Japan
}

\abst{The density-functional (DF) theory provides a simple method for 
calculating the properties of an interacting system under 
an external potential by associating it with a corresponding 
non-interacting system.
Here, we find some relations in 
this non-interacting system, which can enable us to set up the thermodynamical relations for a neutral electron-nucleus mixture in terms of quantities of 
the non-interacting system and the exchange-correlation effect. 
In this way, Andersen's force theorems are clearly described and easily 
proved in conjunction with other types of force theorems, 
Janak's theorem extended to a finite-temperature system, 
pressure formulas and some thermodynamical relations. 
For this purpose, thermodynamics in the DF theory is presented 
in a systematic and explicit way.
}

\begin{document}

\maketitle

\section{Introduction}
The density-functional (DF) theory has been used with success to 
investigate the equilibrium properties of solids
 and liquid metals 
in the almost cases when the electrons can be treated as being in a perfectly degenerate state.
Therefore, not so much attention is paid to thermodynamics in the 
DF theory, although the DF theory is extended to be applicable to 
finite-temperature systems.\cite{KohnSham65,Mermin65}
As a consequence, we found some confusions in applications of the DF theory 
to condensed matter, as shown by some examples below.
Frequently,\cite{R69,W79,Skriver84} the pressure of a solid is determined by 
the derivative of total electronic energy $E_0$ 
with respect to 
the atomic volume $\Omega$ (a uniform compression of the lattice): $P=-dE_0/d\Omega$. 
To be correct, the volume derivative of electronic energy should be 
done under the fixed nuclear positions (fixed lattice) for obtaining the `electron' pressure, as shown in the present work.
For a next example, the proofs of a generalized Janak's theorem to a finite-temperature and the familiar Fermi distribution law are inappropriately proceeded,\cite{Callaway84} since 
the thermodynamic relations in the DF theory are not recognized clearly.
The last example is the entropy expression~\cite{KohnSham65,Rajago80}
 in the DF theory, 
the derivation of which is not transparent.
In this situation, it is useful to write up thermodynamics in the 
DF theory applied to a neutral electron-nucleus mixture 
in the explicit and systematic form.

Andersen\cite{Andersen85} has derived force theorems~\cite{Andersen85,MackAnd85,Heine80,Skriver84} associating a certain change in 
the sum of eigenvalues determined by the DF theory due to 
the volume change or 
atomic displacements with the pressure or the forces, respectively.
However, the expression of force theorems is ambiguous and not definite 
to be used for calculations of the pressure and the forces. 
In addition, their proofs~\cite{Andersen85,MackAnd85,Heine80} are only given on the basis of the physical 
picture as Heine~\cite{Heine80} has done 
even though in long four pages. Furthermore, the force theorems are 
proved 
only at zero temperature and in the local-density approximation (LDA).
Contrary to this situation, thermodynamics in the DF theory 
is shown to provide clear representations of the force theorems extended for 
a finite-temperature system without use of the LDA, 
in conjunction with their simple proofs in the present work.

\section{Thermodynamics of inhomogeneous system \\
under the external potential $U({\bf r})$}

In this section, we sketch a brief outline of thermodynamics of the inhomogeneous 
system for the purpose to set up thermodynamics in the DF theory. 
When the thermodynamic potential $\Omega$ of an inhomogeneous system 
under the external potential $U({\bf r})$ is given, the density distribution 
$n({\bf r})$ is determined by the functional derivative of $\Omega$ 
with respect to $\gamma({\bf r})\equiv \mu-U({\bf r})$ \cite{Chihara78}:
\begin{equation}\label{e:OMn}
\left.\frac{\delta\Omega}{\delta\gamma({\bf r})}\right|_{TV}
=-\left.\frac{\delta\Omega}{\delta U({\bf r})}\right|_{TV\mu}
=-n({\bf r})\,,
\end{equation}
at temperature $T$ of the system with a volume $V$ and chemical potential $\mu$. 
Therefore, the natural variables of the thermodynamic potential $\Omega$ for this 
inhomogeneous system are $T$, $V$, $\gamma({\bf r})$ 
or $T$, $V$, $\mu$ and $U({\bf r})$, and the change in $\Omega$ associated with 
any infinitesimal variation in these natural variables is expressed by 
\begin{eqnarray}
d\Omega &=&-SdT-PdV-\int n({\bf r})\delta\gamma({\bf r})d{\bf r}\label{e:dOM1}\\
&=& -SdT-PdV-Nd\mu+\int n({\bf r})\delta U({\bf r})d{\bf r}\label{e:dOM2}\,.
\end{eqnarray}
The relation (\ref{e:OMn}) enables us to replace an independent variable 
$\gamma({\bf r})$ by $n({\bf r})$ with use of the Legendre transformation of $\Omega$, which introduces the intrinsic free energy $\tilde F$: 
\begin{equation}
{\tilde F}\equiv \Omega-\int\left.\frac{\delta\Omega}{\delta\gamma({\bf r})}\right|_{TV}\gamma({\bf r})d{\bf r}
=\Omega+\int n({\bf r})\gamma({\bf r})d{\bf r}\,. 
\end{equation}
Now, the natural variables of ${\tilde F}$ become 
$T$, $V$, $n({\bf r})$, or $T$, $V$, $N$, $\sigma({\bf r})$, and the variation is given by  
\begin{eqnarray}
d{\tilde F} &=&-SdT-PdV+\int \gamma({\bf r})\delta n({\bf r})d{\bf r}\label{e:dtF1}\\
&=& -SdT-PdV+\left[\mu-\!\int U({\bf r})\sigma({\bf r}) d{\bf r}\right] dN-N\!\!\int U({\bf r})\delta \sigma({\bf r}) d{\bf r}\,. \label{e:dtF2}
\end{eqnarray}
Here, $\sigma({\bf r})$ is the shape factor defined by the relation $n({\bf r})\equiv N\sigma({\bf r})$ \cite{DeProft97}.
\begin{table}[t] 
\caption{State functions and their changes with respect to infinitesimal 
variations in the natural variables.}\label{t:1}
\begin{center}
\renewcommand{\arraystretch}{1.3}
\let\tabularsize\normalsize
\begin{tabular}{|l|l|}
  \hline\hline
state function & natural variables \\ \hline\hline
thermodynamic potential $\Omega$ & $T,V,\gamma({\bf r})$ 
or $T,V,\mu, U({\bf r})$\\ \hline
\multicolumn{2}{|l|}{$d\Omega =-SdT-PdV-\int n({\bf r})\delta\gamma({\bf r})d{\bf r}$\hfill (A1)}\\
\multicolumn{2}{|l|}{\hspace{0.4cm} $=-SdT-PdV-Nd\mu+\int n({\bf r})\delta U({\bf r})d{\bf r}$\hfill (A2)  }\\
\hline\hline
intrinsic thermodynamic potential $\tilde\Omega$ & $T,V,\mu,n({\bf r})$\\ \hline
\multicolumn{2}{|l|}{$d{\tilde\Omega} =-SdT-PdV-Nd\mu-\int U({\bf r})\delta n({\bf r})d{\bf r}$\hfill (B)}\\
\hline\hline
free energy $F$ & $T,V,N,U({\bf r})$\\ \hline
\multicolumn{2}{|l|}{$d{F} =-SdT-PdV+\mu dN+\int n({\bf r})\delta U({\bf r})d{\bf r}$\hfill (C)}\\
\hline\hline
intrinsic free energy $\tilde F$ & $T,V,n({\bf r})$ 
or $T,V,N, \sigma({\bf r})$\\ \hline
\multicolumn{2}{|l|}{$d\tilde{F} =-SdT-PdV+\int \gamma({\bf r})\delta n({\bf r})d{\bf r}$\hfill (D1)}\\
\multicolumn{2}{|l|}{\hspace{-0.1cm} $=-SdT\!-\!PdV\!+[\,\mu\!-\!\!\int\! U({\bf r})\sigma({\bf r})d{\bf r}\,]dN\!-\!N\!\!\int\!U({\bf r})\delta\sigma({\bf r})d{\bf r}$\hfill (D2)  }\\
\hline\hline
internal energy $E$ & $S,V,N,U({\bf r})$\\ \hline
\multicolumn{2}{|l|}{$d{E} =TdS-PdV+\mu dN+\int n({\bf r})\delta U({\bf r})d{\bf r}$\hfill (E)}\\
\hline\hline
intrinsic internal energy $\tilde E$ & $S,V,n({\bf r})$ 
or $S,V,N, \sigma({\bf r})$\\ \hline
\multicolumn{2}{|l|}{$d\tilde E =TdS-PdV+\int \gamma({\bf r})\delta n({\bf r})d{\bf r}$\hfill (F1)}\\
\multicolumn{2}{|l|}{\hspace{-0.1cm} $=TdS\!-\!PdV+[\,\mu\!-\!\!\int\!U({\bf r})\sigma({\bf r})d{\bf r}\,]dN\!-\!N\!\!\int\! U({\bf r})\delta\sigma({\bf r})d{\bf r}$\hfill (F2)  }\\
\hline
\end{tabular}
\renewcommand{\arraystretch}{1.0}
\end{center}
\end{table}

In a similar manner, the free energy $F$, the internal energy $E$, the intrinsic internal energy $\tilde E$ and the intrinsic thermodynamic potential $\tilde \Omega$ for the inhomogeneous system are introduced by means of the following Legendre transforms: 
\begin{eqnarray}
F&\equiv& \Omega-\mu\left.\frac{\partial \Omega}{\partial \mu}\right|_{TVU({\bf r})}
=\Omega+\mu N\,,\\
E&=&F-T\left.\frac{\partial F}{\partial T}\right|_{VNU({\bf r})}=F+TS\,,\\
\tilde E&=&{\tilde F}-T\left.\frac{\partial \tilde F}{\partial T}\right|_{Vn({\bf r})}={\tilde F}+TS\,,\\
{\tilde \Omega}&=&\Omega-\int\left.\frac{\delta\Omega}{\delta U({\bf r})}\right|_{TV\mu}U({\bf r})d{\bf r}
=\Omega-\int n({\bf r})U({\bf r})d{\bf r}\,,
\end{eqnarray}
of $\Omega$, $F$, $\tilde{F}$ and $\Omega$, respectively.
Note that \lq intrinsic' thermodynamical quantities $\tilde Q$ 
(such as ${\tilde E}$, ${\tilde F}$ and 
${\tilde \Omega}$) are related to the usual quantities $Q$ by the relation:
\begin{eqnarray}\label{e:intrinsicQ}
Q={\tilde Q}+\int U({\bf r})n({\bf r})d{\bf r}\,.
\end{eqnarray}
Changes of these thermodynamic quantities associated with 
infinitesimal variations in the natural variables are summarized 
in Table.~\ref{t:1} with additional new relations.

Equations~(\ref{e:dOM1}), (\ref{e:dOM2}), (C) and (E) of Table.~\ref{t:1} lead to formulas to determine 
the density distribution $n({\bf r})$ in several ways as
\begin{equation}\label{e:nr}
n({\bf r})=-\left.\frac{\delta\Omega}{\delta\gamma({\bf r})}\right|_{TV}
=\left.\frac{\delta\Omega}{\delta U({\bf r})}\right|_{TV\mu}
=\left.\frac{\delta F}{\delta U({\bf r})}\right|_{TVN}
=\left.\frac{\delta E}{\delta U({\bf r})}\right|_{SVN}\,.
\end{equation}
Inversely, the external potential $U({\bf r})$ is determined as 
\begin{equation}\label{e:muU}
\left.\frac{\delta{\tilde F}}{\delta n({\bf r})}\right|_{TV}=\left.\frac{\delta{\tilde E}}{\delta n({\bf r})}\right|_{SV}=\mu-U({\bf r})\,,
\end{equation}
by using the relations, (D1) and (F1) of Table~\ref{t:1}. 
Similarly, the chemical potential $\mu$ and the pressure $P$ of the 
inhomogeneous system are calculated by the formulas 
\begin{equation}\label{e:mu}
\mu=\left.\frac{\partial F}{\partial N}\right|_{TVU({\bf r})}
=\left.\frac{\partial E}{\partial N}\right|_{SVU({\bf r})}
=\left.\frac{\delta F}{\delta n({\bf r})}\right|_{TVU({\bf r})}
=\left.\frac{\delta E}{\delta n({\bf r})}\right|_{SVU({\bf r})}\,,
\end{equation}
and
\begin{eqnarray}
P&=&-\left.\frac{\partial F}{\partial V}\right|_{TNU({\bf r})}
=-\left.\frac{\partial E}{\partial V}\right|_{SNU({\bf r})}
=-\left.\frac{\partial \Omega}{\partial V}\right|_{T\gamma({\bf r})}\label{e:Pdev1}\\
&=&-\left.\frac{\partial\tilde F}{\partial V}\right|_{Tn({\bf r})}
=-\left.\frac{\partial\tilde E}{\partial V}\right|_{Sn({\bf r})}
=-\left.\frac{\partial\tilde\Omega}{\partial V}\right|_{T\mu n({\bf r})}\,.\label{e:Pdev2}
\end{eqnarray}
Also, the equilibrium density distribution $n({\bf r})$ satisfies the following condition:
\begin{equation}\label{e:nMIN}
\left.\frac{\delta\Omega}{\delta n({\bf r})}\right|_{TV\mu U({\bf r})}
=\left.\frac{\delta F}{\delta n({\bf r})}\right|_{TVNU({\bf r})}
=\left.\frac{\delta E}{\delta n({\bf r})}\right|_{SVNU({\bf r})}=0\,.
\end{equation}

\section{DF theory of electrons under the external potential\\ 
caused by the fixed nuclei}\label{s:DF} 
Let us consider an electron-nucleus mixture consisting of $N_{\rm n}$ nuclei and 
$N=ZN_{\rm n}$ electrons with an atomic number $Z$. When the nuclei behave as classical 
particles, the nuclei are considered as generating the external potential $U({\bf r})$ for the electrons to form an inhomogeneous electron system:~\cite{Chihara91} 
this external potential for the electrons is the Coulomb potential produced by the nuclei fixed at the coordinates ${\bf R}_\alpha$:
\begin{equation}\label{e:Uext}
U({\bf r})=-\sum_\alpha\frac{Ze^2}{|{\bf r}-{\bf R}_\alpha|}\,.
\end{equation}
In treating this inhomogeneous system, an effective external potential $U_{\rm eff}({\bf r})$ can be defined so as to satisfy 
the condition that the true electron density $n({\bf r}|U)$ under the 
external potential $U({\bf r})$ should be identical with 
the non-interacting electron density distribution $n^0({\bf r}|U_{\rm eff})$ under the external potential $U_{\rm eff}({\bf r})$:
\begin{equation}\label{e:dnsn0n}
n^0({\bf r}|U_{\rm eff})\equiv n({\bf r}|U)\,.
\end{equation}
Here, the non-interacting electron density distribution 
$n^0({\bf r}|U_{\rm eff})$ is determined as  
\begin{equation}\label{e:n0Ueff}
n^0({\bf r}|U_{\rm eff})\equiv \sum_i f_i|\phi_i({\bf r})|^2\,,
\end{equation}
in terms of the wave function $\phi_i({\bf r})$ and eigenvalues $\epsilon_i$, 
which obey the wave equation for a single electron:
\begin{equation}\label{e:KSwe}
\left[\frac{-\hbar^2}{2m}\nabla^2+U_{\rm eff}({\bf r})\right]\phi_i({\bf r})=\epsilon_i\phi_i({\bf r})\,,
\end{equation}
and the Fermi distribution $f_i={1}/\{\exp[\beta(\epsilon_i\!-\!\mu_0)\!+\!1\}$ 
with the chemical potential $\mu_0$ of a non-interacting electron gas.

For the purpose to obtain an explicit expression for the external potential 
$U_{\rm eff}({\bf r})$, let us note the relation for intrinsic free energy 
of non-interacting electrons ${\cal F}_s(\equiv \tilde F_s$):
\begin{equation}\label{e:gamm0}
\left.\frac{\delta{\cal F}_s}{\delta n^0({\bf r})}\right|_{TV}
=\left.\frac{\delta{\cal F}_s}{\delta n({\bf r})}\right|_{TV}
=\mu_0-U_{\rm eff}({\bf r})\equiv\gamma_{\rm eff}({\bf r})\,,
\end{equation}
which is (\ref{e:muU}) rewritten for the non-interacting system.
Here, the intrinsic free energy ${\cal F}_s[n^0]$ of the non-interacting system is written in an explicit form:
\begin{eqnarray}
{\cal F}_s[n^0]&=&-\frac{1}{\beta}\sum_{i=1}^\infty \ln\{1+\exp[\beta(\mu_0-\epsilon_i)]\}-\int n^0({\bf r})U_{\rm eff}({\bf r})d{\bf r}+\mu_0N\label{e:intFs1}\\
&=&T_s[n^0]-TS_s[n^0]\label{e:intFs2}\,,
\end{eqnarray}
where $T_s$ and $S_s$ are the intrinsic internal energy $\tilde E_s$ and the entropy of the non-interacting system, respectively, defined by 
\begin{eqnarray}
T_s[n^0]&\equiv&{\tilde E}_s=E_s
-\int n^0({\bf r})U_{\rm eff}({\bf r})d{\bf r}\label{e:keTs}\\
&=&\sum_if_i\int \phi^*_i({\bf r})\frac{-\hbar^2}{2m}\nabla^2\phi_i({\bf r})d{\bf r}\,,\\
S_s[n^0]&\equiv& -k_{\rm B}\sum_i[f_i\ln f_i+(1-f_i)\ln (1-f_i)]\,\label{e:entSs}
\end{eqnarray}
with the internal energy $E_s=\sum_if_i\epsilon_i$.
On use of the relation:
\begin{equation}\label{e:gamm}
\left.\frac{\delta{\tilde F}}{\delta n({\bf r})}\right|_{TV}
=\mu-U({\bf r})\equiv\gamma({\bf r})\,,
\end{equation}
the effective external potential $U_{\rm eff}({\bf r})$ is represented explicitly as
\begin{equation}\label{e:Ueff1}
U_{\rm eff}({\bf r})=U({\bf r})+\left.\frac{\delta {\cal F}_{\rm I}}{\delta n({\bf r})}\right|_{TV}
-\mu_{\rm I}
\end{equation}
with ${\cal F}_{\rm I}\equiv{\tilde F}-{\tilde F}_s$ and $\mu_{\rm I}\equiv\mu-\mu_0$, which is derived by subtracting (\ref{e:gamm0}) from (\ref{e:gamm}).
Equation (\ref{e:Ueff1}) is the effective external potential in the framework 
of the DF theory. 
At this point, we introduce the exchange-correlation free energy by \begin{equation}\label{e:tFxc}
{\cal F}_{\rm xc}\equiv {\cal F}_{\rm I}-\tilde E_{\rm es}\,,
\end{equation}
in terms of the intrinsic electrostatic energy $\tilde E_{\rm es}$, a part of the electrostatic energy $E_{\rm es}[n]$ described as
\begin{eqnarray}
\hspace{-1cm}E_{\rm es}[n]&\equiv &\frac{e^2}2\!\!\int {
 n({\bf r})n({\bf r'})\over |{\bf r}-{\bf r'}|}d{\bf r}d{\bf r'}
+\frac12\sum_{\alpha\neq\beta}{Z^2 e^2 
\over |{\bf R}_{\alpha}-{\bf R}_{\beta}|} 
-\sum_{\ell=1}^N\!\int {Z e^2 n({\bf r})\over |{\bf r}-
{\bf R}_{\ell}| } d{\bf r}\label{e:Ees}\\
&\equiv &E_{\rm ee}+E_{\rm nn}+E_{\rm en}\equiv\tilde E_{\rm es} +\int n({\bf r})U({\bf r})d{\bf r}\,.\label{e:tEes}
\end{eqnarray}
Then, we can write the intrinsic free energy ${\tilde F}$ in the form:
\begin{equation}\label{e:tFdf}
{\tilde F}={\cal F}_s+{\cal F}_{\rm I}={\cal F}_s+{\cal F}_{\rm xc}+\tilde E_{\rm es}\,,
\end{equation}
within the framework of the DF theory.

With use of the Legendre transformations of $\tilde F$, we obtain the thermodynamic potential in the form:
\begin{eqnarray}
\Omega&=&{\tilde F}-\int \left.\frac{\delta {\tilde F}}{\delta n({\bf r})}\right|_{TV}n({\bf r})d{\bf r}
=F-\mu N\\
&=&\Omega_s+\Omega_{\rm xc}+\Omega_{\rm es}\,,\label{e:OM2df}
\end{eqnarray}
where
\begin{eqnarray}
\Omega_s&=&\frac1\beta \sum_i \ln(1-f_i)=-\frac{1}{\beta}\sum_{i=1}^\infty \ln\{1+\exp[\beta(\mu_0-\epsilon_i)]\}\,,\label{e:OMs2}\\
\Omega_{\rm xc}&=&{\cal F}_{\rm xc}-\int \left.\frac{\delta {\cal F}_{\rm xc}}{\delta n({\bf r})}\right|_{TV}n({\bf r})d{\bf r}\,,\\
\Omega_{\rm es}&=&{\tilde E}_{\rm es}-\int \left.\frac{\delta {\tilde E}_{\rm es}}{\delta n({\bf r})}\right|_{TV}n({\bf r})d{\bf r}=E_{\rm nn}-E_{\rm ee}\,,
\end{eqnarray}
and the intrinsic internal energy ${\tilde E}$ in the form:
\begin{equation}\label{e:tEdf}
{\tilde E}={\tilde F}-T\left.\frac{{\partial \tilde F}}{\partial T}\right|_{Vn({\bf r})}
=T_s[n]+E_{\rm xc}[n]+\tilde E_{\rm es}
\end{equation}
with the exchange-correlation part of the internal energy defined by
\begin{equation}\label{e:Exc}
E_{\rm xc}[n]\equiv {\cal F}_{\rm xc}-T\left.\frac{\partial {\cal F}_{\rm xc}}{\partial T}\right|_{Vn({\bf r})}\,.
\end{equation}
From the above expressions for the intrinsic quantities, $\tilde E$ and $\tilde F$, the relation (\ref{e:intrinsicQ}) provides 
the internal energy $E$:
\begin{equation}\label{e:Edf}
E=\tilde E +\int n({\bf r})U({\bf r})d{\bf r}=T_s+E_{\rm xc}+E_{\rm es}\,,
\end{equation}
and the free energy $F$: 
\begin{eqnarray}
F&=&{\tilde F}+\int n({\bf r})U({\bf r})d{\bf r}=E-TS\\
&=&{\cal F}_s+{\cal F}_{\rm xc}+E_{\rm es}=T_s-TS_s+{\cal F}_{\rm xc}+E_{\rm es}\label{e:Fdf}
\end{eqnarray}
with the definition of the entropy~\cite{KohnSham65,Rajago80} 
\begin{equation}\label{e:Sdf}
S\equiv-\left.\frac{\partial F}{\partial T}\right|_{VNU({\bf r})}=-\left.\frac{\partial {\tilde F}}{\partial T}\right|_{Vn({\bf r})}
=S_s-\left.\frac{\partial {\cal F}_{\rm xc}}{\partial T}\right|_{Vn({\bf r})}\,.
\end{equation}
Equation (\ref{e:Sdf}) is obtained from Eq.~(\ref{e:tFdf}) and the relation: ${\partial\tilde E_{\rm es}}/{\partial T}|_{Vn({\bf r})}=0$; this derivation is different from 
those of the references~\cite{KohnSham65,Rajago80}, which  are 
in a roundabout way.

By use of the relation
\begin{equation}\label{e:muXC}
\mu_{\rm xc}({\bf r}|n)\equiv \left.\frac{\delta {\cal F}_{\rm xc}[n]}{\delta n({\bf r})}\right|_{TV}
=\left.\frac{\delta E_{\rm xc}[n]}{\delta n({\bf r})}\right|_{SV}\,,
\end{equation}
we can express the effective external potential (\ref{e:Ueff1}) as
\begin{equation}\label{e:Ueff2}
U_{\rm eff}({\bf r})=\left.\frac{\delta E_{\rm es}}{\delta n({\bf r})}\right|_{U({\bf r})}+\mu_{\rm xc}({\bf r}|n)-\mu_{\rm I}\,.
\end{equation}
Thus, the effective external potential is rewritten in simple forms:
\begin{equation}\label{e:Ueffdn} 
U_{\rm eff}({\bf r})=\left.\frac{\delta [E_{\rm es}+{\cal F}_{\rm xc}]}{\delta n({\bf r})}\right|_{TVU({\bf r})}-\mu_{\rm I}
=\left.\frac{\delta [E_{\rm es}+{E}_{\rm xc}]}{\delta n({\bf r})}\right|_{SVU({\bf r})}-\mu_{\rm I}\,,
\end{equation}
where we make a convention that the derivative $\delta E_{\rm es}/\delta n({\bf r})$ is performed with the external potential $U({\bf r})$ (that is, the nuclear coordinates $\{{\bf R}_\alpha\}$) being fixed.
\section{Thermodynamics of an ideal gas under the external potential $U_{\rm eff}({\bf r})$}
As shown in the previous section, the DF theory reduces the many-electron problem under the external 
potential to the one-electron problem by associating an interacting system to a non-interacting system (an ideal gas under the effective external potential). 
As a consequence of this association, the thermodynamical properties of 
an interacting system are described in terms of those of the ideal gas
 under the effective external potential.  For this purpose, we enumerate here
some important properties of an ideal Fermi gas.

The change of the intrinsic internal energy $\tilde E_s=T_s[n]$ of the 
non-interacting system is given by
\begin{equation}\label{e:dtEs}
d\tilde E_s=TdS_s-P_0dV+\int \gamma_{\rm eff}({\bf r})\delta n({\bf r})d{\bf r}
=dT_s[n]\,,
\end{equation}
while the change of the intrinsic free energy $\tilde F_s\equiv{\cal F}_s$ is 
\begin{equation}\label{e:dtFs}
d\tilde F_s=-S_sdT-P_0dV+\int \gamma_{\rm eff}({\bf r})\delta n({\bf r})d{\bf r}
=d{\cal F}_s[n]\,.
\end{equation}
Therefore, from the above equations, the external potential $\gamma_{\rm eff}({\bf r})=\mu_0-U_{\rm eff}({\bf r})$
is obtained by the functional derivatives in the two ways:
\begin{equation}\label{e:mu0Ueff}
\left.\frac{\delta{\cal F}_s[n]}{\delta n({\bf r})}\right|_{TV}
=\left.\frac{\delta T_s[n]}{\delta n({\bf r})}\right|_{S_sV}
=\mu_0-U_{\rm eff}({\bf r})\equiv\gamma_{\rm eff}({\bf r})\,,
\end{equation}
which leads to (\ref{e:muXC}) with aids of (\ref{e:muU}). 
Also, the pressure $P_0$ 
is derived from (\ref{e:dtEs}) and (\ref{e:dtFs}) as
\begin{eqnarray}
3P_0V&=&-3V\left.\frac{\partial T_s[n]}{\partial V}\right|_{S_sn({\bf r})}
=-3V\left.\frac{\partial {\cal F}_s[n]}{\partial V}\right|_{Tn({\bf r})}\label{e:p0V1}\\
&=&2T_s[n]-\int n({\bf r}){\bf r}\cdot\nabla U_{\rm eff}({\bf r})d{\bf r}\label{e:p0V2}\\
&=&\oint_{\partial V}{\bf r}\cdot{\bf\textsf P}_{\rm K}^0\cdot d{\bf S}\,,\label{e:p0V3}
\end{eqnarray} 
where the kinetic pressure tensor ${\bf\textsf P}_{\rm K}^0$ is defined 
by~\cite{Chihara01}
\begin{equation}\label{e:Pk0Def}
\nabla\cdot{\bf\textsf P}_{\rm K}^0
=-n({\bf r})\nabla U_{\rm eff}({\bf r})\,.
\end{equation}
The above result is derived as follows. 
First, with use of the scaled density $n_\lambda({\bf r})\equiv \lambda^3n(\lambda{\bf r})$,\cite{LevyPer85} the kinetic energy functional is shown to 
satisfy the relation: $T_s[n_\lambda]=\lambda^2T_s[n]$, which yields 
\begin{equation}\label{e:dVTs}
-3V\left.\frac{\partial T_s[n]}{\partial V}\right|_{S_sNU_{\rm eff}({\bf r})}
=\left.\lambda{dT_s[n_\lambda] \over d\lambda}\right|_{\lambda=1}=2T_s[n]\,.
\end{equation}
Next, we note that the volume derivative of $T_s$ at the fixed $S_s,N,U_{\rm eff}$ 
is expressed by a sum of the volume derivatives of $T_s$ 
at the fix $n({\bf r})$ and via the change in $n({\bf r})$:
\begin{equation}\label{e:dTsdV}
\left.\frac{\partial T_s[n]}{\partial V}\right|_{S_sNU_{\rm eff}({\bf r})}
=\left.\frac{\partial T_s[n]}{\partial V}\right|_{S_sn({\bf r})}
+\int \left.\frac{\delta T_s[n]}{\delta n({\bf r})}\right|_{S_sV}
\left.\frac{dn({\bf r})}{dV}\right|_{S_sNU_{\rm eff}}\!\!\!d{\bf r}\,,
\end{equation}
(see Appendix).
Here, the first term in the right side of (\ref{e:dTsdV}) is expressed as  
\begin{equation}\label{e:dTsdV2}
-3V\left.\frac{\partial T_s[n]}{\partial V}\right|_{S_sn({\bf r})}
=2T_s[n]+\int n({\bf r}){\bf r}\cdot\nabla \left.\frac{\delta T_s[n]}{\delta n({\bf r})}\right|_{S_sV}d{\bf r}\,,
\end{equation}
with help of (\ref{e:dVTs}) and the following equation:~\cite{Abbreb1}
\begin{eqnarray}
-3V&&\int f({\bf r})\frac{dn({\bf r})}{dV}d{\bf r}
=\left.\lambda\frac{d}{d\lambda}\int_{V/\lambda^3}f({\bf r})n_\lambda({\bf r})d{\bf r}\right|_{\lambda=1}\nonumber\\
&&=\lambda\frac{d}{d\lambda}\int_{V}f({\bf r}/\lambda)n({\bf r})d{\bf r}
=-\int_V n({\bf r}){\bf r}\cdot\nabla f({\bf r})d{\bf r}\,,\label{e:fdndV}
\end{eqnarray}
which is applied to the second term in the right side of (\ref{e:dTsdV}).
Finally, as a consequence of (\ref{e:dTsdV2}) and (\ref{e:mu0Ueff}), we obtain (\ref{e:p0V2}), which may be written in the surface integral (\ref{e:p0V3}) over the surface $\partial V$ of the system.~\cite{Chihara01} 
It is important to recognize that the kinetic energy of a non-interacting electron gas is written as
\begin{eqnarray}\label{e:TsTr}
2T_s[n]=\int_V n({\bf r}){\bf r}\cdot\nabla U_{\rm eff}({\bf r})d{\bf r}+\oint_{\partial V}{\bf r}\cdot{\bf\textsf P}_{\rm K}^0\cdot d{\bf S}
=\int_V {\rm tr} {\bf\textsf P}_{\rm K}^0 d{\bf r}\,,
\end{eqnarray} 
which yields the surface-integral expression (\ref{e:p0V3}). The surface integral term 
is omitted in the standard treatment \cite{Slater72,Janak74,Rajago80}, which yields erroneous results if we treat a system with the non-zero electron pressure 
by using (\ref{e:TsTr}) without the surface integral term. For example, 
Slater\cite{Slater72} derived the virial theorem, 2(kinetic energy)+(potential 
energy)$=-\sum_\alpha {\bf R}_\alpha \nabla_\alpha E$, which is valid only for 
a solid with the zero electron-pressure, as will be shown afterward by (\ref{e:PeVir}) and (\ref{e:PeVir2}).

When we differentiate the internal energy $E_s=\sum_if_i\epsilon_i$ in a simple manner, we find
\begin{equation} 
\frac{d}{d f_i}\sum_if_i\epsilon_i
=\epsilon_i
+f_i\frac{d \epsilon_i}{d f_i}\,.
\end{equation}
It is important to see the reason why the second term is discarded to yield 
$\epsilon_i$ in the circumstance that $f_i$ is a function of $\epsilon_i$.
As is seen later, this reasoning provides a clue to relating thermodynamics in the DF theory to usual thermodynamics. First, we consider the derivative of $E_s$ with respect to an arbitrary parameter $\lambda$ by noting the relation, $E_s=\sum_if_i\epsilon_i=\Omega_s+TS_s+\mu_0N$.
Because of ${\partial\Omega_s}/{\partial T}|_{V\mu_0U_{\rm eff}}=-S_s$, we get the following relation:
\begin{eqnarray}
\frac{d}{d\lambda}[\Omega_s+TS_s]&=&\left.\frac{d\Omega_s}{d\lambda}\right|_T
+\left[\left.\frac{\partial\Omega_s}{\partial T}\right|_{V\mu_0U_{\rm eff}}+S_s\right]\frac{dT}{d\lambda}+T\frac{dS_s}{d\lambda}\\
&=&\left.\frac{d\Omega_s}{d\lambda}\right|_T+T\frac{dS_s}{d\lambda}\,, \label{e:dlamOsSs}
\end{eqnarray}
where the first term of (\ref{e:dlamOsSs}) is put in the form
\begin{equation}
\left.\frac{d\Omega_s}{d\lambda}\right|_T
=\left.\frac{d\Omega_s}{d\lambda}\right|_{T\mu_0}
+\left.\frac{\partial\Omega_s}{\partial \mu_0}\right|_{TVU_{\rm eff}}\!\!\!\!\frac{d\mu_0}{d\lambda}
=\left.\frac{d\Omega_s}{d\lambda}\right|_{T\mu_0}
\!\!\!\!-N\frac{d\mu_0}{d\lambda}\,.\label{e:dOmdlamT}
\end{equation}
Thus, from (\ref{e:dlamOsSs}) and (\ref{e:dOmdlamT}) we obtain a simple expression (see Appendix)
\begin{equation}\label{e:dEsdlam}
\frac{dE_s}{d\lambda}=\left.\frac{d\Omega_s}
{d \lambda}\right|_{T\mu_0}+T\frac{dS_s}{d\lambda}
+\mu_0\frac{dN}{d\lambda}\,.
\end{equation}
If we take as $\lambda=f_i$, the above equation takes the form
\begin{equation}\label{e:dEsdfi}
\left.\frac{dE_s}{df_i}\right|_{VU_{\rm eff}}\!\!=\left.\frac{\partial\Omega_s}
{\partial f_i}\right|_{TV\mu_0U_{\rm eff}}\!\!-\frac1{\beta}\ln\frac{f_i}{1-f_i}
+\mu_0
=\left.\frac{\partial\Omega_s}{\partial f_i}\right|_{TV\mu_0U_{\rm eff}}
\!\!+\epsilon_i\,,
\end{equation}
which is consistent only with the following three equations:
\begin{eqnarray}
&&f_i=f(\epsilon_i-\mu_0)\equiv\frac1{\exp[\beta(\epsilon_i-\mu_0)]+1}\label{e:fermD2}\,,\\
&&\left.\frac{d}{d f_i}\left[\sum_if_i\epsilon_i\right]\right|_{VU_{\rm eff}}
=\epsilon_i\label{e:dfiei}\,,\\
&&\left.\frac{\partial\Omega_s}{\partial f_i}\right|_{TV\mu_0U_{\rm eff}}=0\,.\label{e:dOMdfi}
\end{eqnarray}

By taking account of this situation (\ref{e:dEsdlam}), we set up a formula to perform 
a derivative of intrinsic free energy ${\tilde E}_s=T_s[n]$ by $\lambda$, 
which is not contained in $U_{\rm eff}$, in the form:
\begin{equation}\label{e:dTsdlam}
\frac{dT_s}{d\lambda}=\left.\frac{d\Omega_s}{d\lambda}\right|_{T\mu_0 U_{\rm eff}}+T\frac{dS_s}{d\lambda}+
\int [\mu_0-U_{\rm eff}({\bf r})]\frac{dn({\bf r})}{d\lambda}d{\bf r}\,.
\end{equation}
This equation plays an important role to associate the DF theory with 
thermodynamics of an inhomogeneous electron system caused by 
the external potential of the fixed nuclei.
The proof of this equation is shown as follows. 
The kinetic energy functional $T_s$ is written in the form
\begin{equation}\label{e:OsSsng}
T_s=E_s-\int n({\bf r})U_{\rm eff}({\bf r})d{\bf r}
=\Omega_s+TS_s+\int n({\bf r})\gamma_{\rm eff}({\bf r})d{\bf r}\,,
\end{equation}
and Eq.~(\ref{e:dlamOsSs}) is written as
\begin{eqnarray}
\frac{d}{d\lambda}[\Omega_s+TS_s]
=\left.\frac{d\Omega_s}{d\lambda}\right|_T
+T\frac{dS_s}{d\lambda}=\sum_if_i\frac{d[\epsilon_i\!-\!\mu_0]}{d\lambda}
+\sum_i\frac{df_i}{d\lambda}[\epsilon_i\!-\!\mu_0]\,.\label{e:dOmTS}
\end{eqnarray}
Here, the second term of (\ref{e:dOmTS}) is written as
\begin{eqnarray}
\left.\frac{d\Omega_s}{d\lambda}\right|_{T}
&=&\left.\frac{d\Omega_s}{d\lambda}\right|_{T\gamma_{\rm eff}}
\!\!+\int\left.\frac{\delta\Omega_s}{\delta\gamma_{\rm eff}({\bf r})}
\right|_{TV}\!\!\frac{\delta \gamma_{\rm eff}({\bf r}) }{\delta n({\bf r'})}
\frac{dn({\bf r'})}{d\lambda}d{\bf r}d{\bf r'}\\
&=&\left.\frac{d\Omega_s}{d\lambda}\right|_{T\mu_0U_{\rm eff}}\!\!
-\int n({\bf r})\frac{\delta \gamma_{\rm eff}({\bf r}) }{\delta n({\bf r'})}
\frac{dn({\bf r'})}{d\lambda}d{\bf r}d{\bf r'}\,.\label{e:dOsdlamT2}
\end{eqnarray}
Also, the third term of the right side of (\ref{e:OsSsng}) generates 
\begin{equation}\label{e:ngamdlam}
\frac{d}{d\lambda}\int n({\bf r})\gamma_{\rm eff}({\bf r})d{\bf r}
=\int \gamma_{\rm eff}({\bf r})\frac{dn({\bf r})}{d\lambda}d{\bf r}
 +\int n({\bf r})
\frac{\delta \gamma_{\rm eff}({\bf r}) }{\delta n({\bf r'})}
\frac{dn({\bf r'})}{d\lambda}d{\bf r}d{\bf r'}\,.
\end{equation}
In final, a cancellation between the second terms in the right side of (\ref{e:dOsdlamT2}) 
and (\ref{e:ngamdlam}) results in (\ref{e:dTsdlam}).

\section{Force theorems}
In this section, we derive three types of Andersen's force theorems 
on the basis of (\ref{e:dTsdlam}) and (\ref{e:dTsdlamUeff}). 
The first type is a force theorem concerning the pressure. 
The pressure of a inhomogeneous electron system under the external 
potential caused by the fixed nuclei at $\{{\bf R}_\alpha\}$ is defined by (\ref{e:Pdev1}), and can be calculated by introducing the scaled electron density distribution $n_\lambda({\bf r})\equiv \lambda^3n(\lambda{\bf r})$ in the following form:
\begin{equation}\label{e:dEdlam} 
3P_eV=-3V\left.\frac{\partial E^V[n;\{{\bf R}_\alpha\}]}{\partial V}\right|_{SN\{{\bf R}_\alpha\}}
=\left.\lambda \frac{dE_\lambda}{d\lambda}\right|_{\lambda=1}\,, 
\end{equation}
with 
\begin{equation}\label{e:Elam}
E_\lambda\equiv E^{V/\lambda^3}[n_\lambda;\{{\bf R}_\alpha\}]
=T_s^{V/\lambda^3}[n_\lambda]+E_{\rm es}^{V/\lambda^3}[n_\lambda;\{{\bf R}_\alpha\}]+E_{\rm xc}^{V/\lambda^3}[n_\lambda]\,.
\end{equation}
With aids of (\ref{e:dTsdlam}), we get~ \cite{Abbreb1} 
\begin{eqnarray}
\left.\frac{dT_s}{d\lambda}\right|_{SN\{{\bf R}_\alpha\}}&=&\left.\frac{d\Omega_s}{d\lambda}\right|_{T\mu_0 U_{\rm eff}}+\frac{d}{d\lambda}\int_{V/\lambda^3} n_\lambda({\bf r})[\mu_0-U_{\rm eff}({\bf r})]d{\bf r}\\
&=&\sum_i\left.f_i\frac{d\epsilon_i}{d\lambda}\right|_{U_{\rm eff}}+\int_V n({\bf r}){\bf r}\cdot\nabla U_{\rm eff}({\bf r})d{\bf r}\,.\label{e:dTsdep}
\end{eqnarray}
Since the electrostatic energy~(\ref{e:tEes}) satisfies the following relation
\begin{equation}\label{e:Eeslam}
E_{\rm es}^{V/\lambda^3}[n_\lambda;\{{\bf R}_\alpha\}]
=\lambda E_{\rm es}^V[n;\{\lambda {\bf R}_\alpha\}]\,,
\end{equation}
the $\lambda$-derivative of the electrostatic energy~\cite{Chihara01} becomes  
\begin{equation}\label{e:dEesdlam}
\left.\lambda\frac{d E_{\rm es}^{V/\lambda^3}[n_\lambda]}{d \lambda}\right|_{\lambda=1}\!\!=E_{\rm es}
-\sum {\bf R}_{\alpha}\!\cdot\!{\bf F}_\alpha
=-\int_V n({\bf r}){\bf r}\cdot\nabla{\delta {E}_{\rm es}\over\delta n({\bf r})}d{\bf r}\,,
\end{equation}
where the force on $\alpha$-nucleus is given by ${\bf F}_\alpha\equiv -\nabla_\alpha E_{\rm es}$\,.
Also, the exchange-correlation energy part \cite{Chihara01} of (\ref{e:Elam}) yields 
\begin{equation}\label{e:dExcdlam}
\left.\lambda \frac{d E_{\rm xc}^{V/\lambda^3}[n_\lambda]}{d \lambda}\right|_{\lambda=1}\!\!=\int_V {\rm tr}{\bf\textsf P}_{\rm xc}d{\bf r}=-\int_V n({\bf r}){\bf r}\cdot\nabla{\delta {E}_{\rm xc}\over\delta n({\bf r})} d{\bf r}+\oint_{\partial V}{\bf r}\cdot{\bf\textsf P}_{\rm xc}\cdot d{\bf S}
\,,
\end{equation}
where the exchange-correlation pressure tensor ${\bf\textsf P}_{\rm xc}$ 
is defined by
\begin{equation}\label{e:PxcDef}
\nabla\cdot {\bf\textsf P}_{\rm xc}\equiv n({\bf r})\nabla 
\left.{\delta {\cal F}_{\rm xc}\over\delta n({\bf r})}\right|_{TV}
=n({\bf r})\nabla\left.{{\delta E}_{\rm xc}\over\delta n({\bf r})}\right|_{SV}\,.
\end{equation} 
Because of (\ref{e:Ueffdn}), 
the sum of the second term in (\ref{e:dTsdep}), 
the last term of (\ref{e:dEesdlam}) and the first term 
in the right side of (\ref{e:dExcdlam}) becomes zero owing to (\ref{e:Ueffdn}); as a result, a pressure formula (the force theorem concerning the electron pressure \cite{Skriver84}) is written as 
\begin{equation}\label{e:forceT1}
3P_eV
=-3V\sum_i\left.f_i\frac{d\epsilon_i}{dV}\right|_{U_{\rm eff}}
+\oint_{\partial V}{\bf r}\cdot{\bf\textsf P}_{\rm xc}\cdot d{\bf S}\,,
\end{equation}
in terms of eigenvalues $\epsilon_i$ of the Kohn-Sham equation (\ref{e:KSwe}).

Another important relation is derived by noting the relation~\cite{LevyPer85}
\begin{equation}\label{e:Tslam}
T_s^{V/\lambda^3}[n_\lambda]=\lambda^2 T_s^V[n]\,,
\end{equation}
which leads to
\begin{eqnarray}
\left.\frac{dT_s}{d\lambda}\right|_{SN\{{\bf R}_\alpha\}}\!\!\!\!&=&2T_s[n]
=\sum_i\left.f_i\frac{d\epsilon_i}{d\lambda}\right|_{U_{\rm eff}}+\int_V n({\bf r}){\bf r}\cdot\nabla U_{\rm eff}({\bf r})d{\bf r}\label{e:dTsSNU}\\
&=&\int_V {\rm tr}{\bf\textsf P}^0_{\rm K}d{\bf r}=\oint_{\partial V}{\bf r}\cdot{\bf\textsf P}^0_{\rm K}\cdot d{\bf S}+\int_V n({\bf r}){\bf r}\cdot\nabla U_{\rm eff}({\bf r}) d{\bf r}\,,\label{e:trPk}
\end{eqnarray}
with combined use of (\ref{e:dTsdep}). 
From the above equation, we relate the volume derivative of eigenvalues $\epsilon_i$ to the surface integral of the pressure tensor ${\bf\textsf P}^0$ for a non-interacting electron gas:
\begin{equation}\label{e:PkSurf}
-3V\sum_i\left.f_i\frac{d\epsilon_i}{dV}\right|_{U_{\rm eff}}
=\oint_{\partial V}{\bf r}\cdot{\bf\textsf P}^0_{\rm K}\cdot d{\bf S}=3P_0V\,,
\end{equation}
which alters a pressure formula (\ref{e:forceT1}) in the form
\begin{equation}\label{e:PeSurf}
3P_eV
=\oint_{\partial V}{\bf r}\cdot({\bf\textsf P}^0_{\rm K}+{\bf\textsf P}_{\rm xc})\cdot d{\bf S}\,.
\end{equation}
At this point, we remark that Eq.~(\ref{e:PkSurf}) is directly derived 
as the pressure of the ideal gas from (\ref{e:p0V3}) and the thermodynamic relation:
\begin{equation}
3P_0V=-3V\left.\frac{\partial\Omega_s}{\partial V}\right|_{T\mu_0 U_{\rm eff}}
=-3V\sum_i\left.f_i\frac{d\epsilon_i}{dV}\right|_{U_{\rm eff}}\,.
\end{equation}

The formula (\ref{e:dTsdlam}) to calculate the $\lambda$-derivative of 
$T_s$ is valid only when the parameter $\lambda$ is not contained in the 
external potential $U_{\rm eff}$. When the parameter $\lambda$ is involved in 
the external potential $U_{\rm eff}$ (for example, $\lambda={\bf R}_\alpha$), 
the formula (\ref{e:dTsdlam}) is changed in the form:
\begin{equation}\label{e:dTsdlamUeff}
\frac{dT_s}{d\lambda}=\left.\frac{d\Omega_s}{d\lambda}\right|_{T\mu_0 n({\bf r})}\!\!\!\!+T\frac{dS_s}{d\lambda}
+\int \gamma_{\rm eff}({\bf r})\frac{dn({\bf r})}{d\lambda}d{\bf r}
-\int \!n({\bf r})\left.\frac{dU_{\rm eff}({\bf r})}{d\lambda}\right|_{n({\bf r})}\!\!d{\bf r}\,,
\end{equation} 
where the derivative of $\Omega_s$ is performed under 
the fixed density distribution $n({\bf r})$, 
instead of the external potential $U_{\rm eff}$ being fixed as (\ref{e:dTsdlam}), 
and the last term of (\ref{e:dTsdlamUeff}) contains the direct variation of $U_{\rm eff}$. 
Actually, Eq.~(\ref{e:dTsdlamUeff}) is derived in a similar way to get 
Eq.~(\ref{e:dTsdlam}) with use of the fact that $T_s=\tilde\Omega_s+TS_s+\mu_0N$ 
and that the natural variables of the intrinsic thermodynamic potential $\tilde\Omega_s$ are $T, V, \mu_0$ and $n({\bf r})$. 
When the $\lambda$-derivative is performed under the fixed $S_s$ and $V$, Eq.~(\ref{e:dTsdlamUeff}) becomes
\begin{equation}\label{e:dTsdlam3}
\left.\frac{dT_s}{d\lambda}\right|_{S_sV}
=\sum_i\left.f_i\frac{d\epsilon_i}{d\lambda}\right|_{n({\bf r})}
+\int \gamma_{\rm eff}({\bf r})\frac{dn({\bf r})}{d\lambda}d{\bf r}
-\int \!n({\bf r})\left.\frac{dU_{\rm eff}({\bf r})}{d\lambda}\right|_{n({\bf r})}\!\!d{\bf r}\,.
\end{equation}
On the other hand, this derivative can be represented in another form:
\begin{equation}\label{e:dTsdlam4}
\left.\frac{dT_s}{d\lambda}\right|_{S_sV}
=\int \left.\frac{\delta T_s[n]}{\delta n({\bf r})}\right|_{S_sV}\frac{dn({\bf r})}{d\lambda}d{\bf r}
=\int \gamma_{\rm eff}({\bf r})\frac{dn({\bf r})}{d\lambda}d{\bf r}\,,
\end{equation}
owing to (\ref{e:mu0Ueff}). 
By comparing (\ref{e:dTsdlam4}) with (\ref{e:dTsdlam3}), we obtain a 
formula to calculate the force on the atoms in the form:
\begin{equation}\label{e:forceTh}
\sum_i\left.f_i\frac{d\epsilon_i}{d\lambda}\right|_{n({\bf r})}
=\int \!n({\bf r})\left.\frac{dU_{\rm eff}({\bf r})}{d\lambda}\right|_{n({\bf r})}\!\!d{\bf r}\,.
\end{equation}
When $\lambda={\bf R}_\alpha$, this equation assumes the form: 
\begin{equation}\label{e:depdR1}
\sum_i\left.f_i\frac{d\epsilon_i}{d{\bf R}_\alpha}\right|_{n({\bf r})}
=\int n({\bf r})\frac{\partial}{\partial {\bf R}_\alpha}\sum_\gamma\frac{-Ze^2}{|{\bf r}-{\bf R}_\gamma|}d{\bf r}
=-Ze^2\int n({\bf r})\frac{{\bf r}-{\bf R}_\alpha}{|{\bf r}-{\bf R}_\alpha|^3}d{\bf r}\,.
\end{equation}
Therefore, the force on $\alpha$-atom is represented as
\begin{eqnarray}
{\bf F}_\alpha&=&-\left.\nabla_\alpha E_{\rm es}\right|_{n({\bf r})}\\
&=&Ze^2\int n({\bf r})\frac{{\bf r}-{\bf R}_\alpha}{|{\bf r}-{\bf R}_\alpha|^3}d{\bf r}
+(Ze)^2\sum_{\alpha\ne \gamma}\frac{{\bf R}_\alpha-{\bf R}_\gamma}{|{\bf R}_\alpha-{\bf R}_\gamma|^3}\label{e:fHF}\\
&=&-\sum_i\left.f_i\frac{d\epsilon_i}{d{\bf R}_\alpha}\right|_{n({\bf r})}
+(Ze)^2\sum_{\alpha\ne \gamma}\frac{{\bf R}_\alpha-{\bf R}_\gamma}{|{\bf R}_\alpha-{\bf R}_\gamma|^3}\,.\label{e:forceT2}
\end{eqnarray}
If we expand the eigenfunction $\phi_i$ by using a local base set $u_i({\bf r})$ as 
$\phi_i({\bf r})=\sum_ju_j({\bf r})c^j_i$, the force is described by the matrix form as is given by
\begin{equation}
\sum_i\left.f_i\frac{d\epsilon_i}{d{\bf R}_\alpha}\right|_{n({\bf r})}
=\sum_i\left(C_i\right|\frac{\partial(H-\epsilon_iS)}{\partial{\bf R}_\alpha}\left|C_i\right)\,,
\end{equation}
with $H_{ij}\equiv (u_i|\hat H|u_j)$ and $S_{ij}\equiv (u_i|u_j)$.\cite{ForceMat}

Next, we consider another type of force theorem \cite{MackAnd85,Heine80}, when not only the nucleus at ${\bf R}_\alpha$, but also the electronic charge within an arbitrary volume 
$\Omega_\alpha$ enclosing the $\alpha$-nucleus is displaced rigidly. 
Instead of displacing the volume $\Omega_\alpha$, it is more convenient 
to think in terms of the displacement of the rest system by $-\delta{\bf R}_\alpha$. Therefore, Eq.~(\ref{e:forceTh}) assumes the form
\begin{eqnarray}
\sum_i\left.f_i\frac{\delta\epsilon_i}{\delta{\bf R}_\alpha}\right|_{n({\bf r})}
&=&\int_{\Omega_\alpha} \!n({\bf r})\frac{\delta[U_{\rm eff}({\bf r}-\delta{\bf R}_\alpha)-U_{\rm eff}({\bf r})]}{\delta{\bf R}_\alpha}d{\bf r}\\
&=&-\int_{\Omega_\alpha} \!n({\bf r})\nabla U_{\rm eff}({\bf r})d{\bf r}\,. \label{e:depsiUeff}
\end{eqnarray} 
On the other hand, the Maxwell tensor ${\bf\textsf T}_{\rm M}$ obeys 
the following equation~\cite{Chihara01}
\begin{equation}\label{e:MaxF}
-\nabla \cdot ({\bf\textsf P}_{\rm K}^{0}
+{\bf\textsf P}_{\rm xc}-{\bf\textsf T}_{\rm M})
=\sum_\alpha{\bf F}_\alpha\delta ({\bf r}-{\bf R}_\alpha)\,. 
\end{equation}
Thus, integral of (\ref{e:MaxF}) over the volume ${\Omega_\alpha}$ yields the force 
on $\alpha$-nucleus in the form:
\begin{eqnarray}
{\bf F}_\alpha&=&-\oint_{\partial \Omega_\alpha}({\bf\textsf P}^0_{\rm K}+
{\bf\textsf P}_{\rm xc}-{\bf\textsf T}_{\rm M})\cdot d{\bf S}\label{e:Fsurf1}\\
&=&-\int_{\Omega_\alpha} \nabla\cdot({\bf\textsf P}^0_{\rm K}+
{\bf\textsf P}_{\rm xc})d{\bf r}
+\oint_{\partial \Omega_\alpha}{\bf\textsf T}_{\rm M}\cdot d{\bf S}\label{e:Fsurf2}\\
&=&\int_{\Omega_\alpha} \!n({\bf r})\nabla U_{\rm eff}({\bf r})d{\bf r}
-\int_{\Omega_\alpha} \!n({\bf r})\nabla \mu_{\rm xc}({\bf r})d{\bf r}
+\oint_{\partial \Omega_\alpha}{\bf\textsf T}_{\rm M}\cdot d{\bf S}\label{e:Fsurf3}\\
&=&-\sum_i\left.f_i\frac{\delta\epsilon_i}{\delta{\bf R}_\alpha}\right|_{n({\bf r})}
-\int_{\Omega_\alpha} \!n({\bf r})\nabla \mu_{\rm xc}({\bf r})d{\bf r}
+\oint_{\partial \Omega_\alpha}{\bf\textsf T}_{\rm M}\cdot d{\bf S}\label{e:forceT3}\,, \label{e:AFT2}
\end{eqnarray}
where the force is described in terms of eigenvalues $\epsilon_i$ due to (\ref{e:depsiUeff}). Equation (\ref{e:AFT2}) should be compared with (193) of the reference~\cite{Andersen85}, where no exchange-correlation contribution is 
contained and the electrostatic force is differently defined.

\section{Thermodynamical relations in DF theory}
In previous section, we have proved Andersen's force theorems on the basis 
of (\ref{e:dTsdlam}) and (\ref{e:dTsdlamUeff}). In this section, we show that 
these equations, (\ref{e:dTsdlam}) and (\ref{e:dTsdlamUeff}), provide several thermodynamical relations in the DF theory. First, we show that the force ${\bf F}_\alpha$ on $\alpha$-nucleus in the system is described by the thermodynamic functions
\begin{equation}\label{e:forceDE}
{\bf F}_\alpha=-\nabla_{\alpha}\left.E\right|_{SVN}
=-\nabla_{\alpha}\left.F\right|_{TVN}
=-\nabla_{\alpha}\left.\Omega\right|_{TV\mu}\,.
\end{equation}
For example, the relation, ${\bf F}_\alpha=-\nabla_{\alpha}[T_s+E_{\rm es}+E_{\rm xc}]|_{SVN}$, is proved by use of the following equations:
\begin{equation}\label{e:DEesxc}
\nabla_{\alpha}\left.[E_{\rm es}+E_{\rm xc}]\right|_{SVN}
=\nabla_{\alpha}\left.E_{\rm es}\right|_{n({\bf r})}+\int \left.\frac{\delta[E_{\rm es}+E_{\rm xc}]}{\delta n({\bf r})}\right|_{SVU({\bf r})}\nabla_{\alpha} n({\bf r})d{\bf r}\,,
\end{equation}
\begin{equation}\label{e:DTsUeff}
\nabla_{\alpha}\left.T_s\right|_{SVN}
=\int \left.\frac{\delta T_{\rm s}}{\delta n({\bf r})}\right|_{S_sV}\nabla_{\alpha} n({\bf r})d{\bf r}=-\int [U_{\rm eff}({\bf r})+\mu_{\rm I}]\nabla_{\alpha} n({\bf r})d{\bf r}\,,
\end{equation}
which lead to ${\bf F}_\alpha=-\nabla_{\alpha}E_{\rm es}|_{n({\bf r})}$ with help of (\ref{e:Ueffdn}).
In the above, we used an important relation to cancel the term involving $U_{\rm eff}$ 
in (\ref{e:dTsdlam}) and (\ref{e:dTsdlamUeff}) 
\begin{equation}\label{e:dnUeff}
\frac{d[E_{\rm es}+E_{\rm xc}]}{d\lambda}
=\left.\frac{d[\tilde E_{\rm es}+E_{\rm xc}]}{d\lambda}\right|_{n({\bf r})}
\!+\!\int \frac{dU({\bf r})}{d\lambda}n({\bf r})d{\bf r} 
\!+\!\int \left.\frac{\delta[E_{\rm es}+E_{\rm xc}]}{\delta n({\bf r})}
\right|_{SVU({\bf r})}\!\!\frac{dn({\bf r})}{d\lambda}d{\bf r}\,,
\end{equation}
which is derived by applying (\ref{e:diffParm}) to the intrinsic internal energy $\tilde E\!=\!E\!-\!\int U({\bf r})n({\bf r})d{\bf r}$ with the natural variables, $S,V$ and $n({\bf r})$.
At this point, it should be noticed that the relation (\ref{e:DTsUeff}) is equivalent to the following equation
\begin{eqnarray}
&&\left.\nabla_\alpha\Omega_s\right|_{TV\mu_0 n({\bf r})}\!
-\!\int \!n({\bf r})\left.\nabla_\alpha U_{\rm eff}({\bf r})\right|_{n({\bf r})}\!\!d{\bf r}=\left.\nabla_\alpha\tilde\Omega_s\right|_{TV\mu_0 n({\bf r})}\\
&&=\sum_i\left.f_i\frac{d\epsilon_i}{d{\bf R}_\alpha}\right|_{n({\bf r})}
+Ze^2\int n({\bf r})\frac{{\bf r}-{\bf R}_\alpha}{|{\bf r}-{\bf R}_\alpha|^3}d{\bf r}=0\,, \label{e:depdR2}
\end{eqnarray}
which results from (\ref{e:dTsdlamUeff}).
Therefore, the force expressions (\ref{e:forceDE}) are 
equivalent to (\ref{e:depdR2}), that is, the force theorem (\ref{e:forceT2}). 
Similarly, other relations in (\ref{e:forceDE}) can be proved with use of 
(\ref{e:dTsdlamUeff}).

Equations (\ref{e:Pdev1}) and (\ref{e:Pdev2}) provide 
two different but equivalent expressions for the pressure as shown below.
First equation (\ref{e:Pdev1}) generates a pressure formula in the form:
\begin{eqnarray}
3P_eV&=&-3V\left.\frac{\partial F}{\partial V}\right|_{TN\{{\bf R}_\alpha\}}
=-3V\left.\frac{\partial E}{\partial V}\right|_{SN\{{\bf R}_\alpha\}}\label{e:Pe1}\\
&=&2T_s[n]+E_{\rm es}+\int {\rm tr} {\bf\textsf P}_{\rm xc}d{\bf r}
-\sum_\alpha {\bf R}_{\alpha}\!\cdot\!{\bf F}_\alpha\,. \label{e:PeVir}
\end{eqnarray}
Equation (\ref{e:PeVir}) is proved from 
(\ref{e:dEesdlam}) and (\ref{e:dExcdlam}) with additional use of (\ref{e:dTsSNU}). Here, it should be kept in mind that the electron pressure is determined by the derivative with the nuclear positions (that is, the external potential $U({\bf r})$) being fixed . In some pressure calculations,~\cite{R69,W79,Skriver84} 
 this fixed-nuclei condition is neglected; as a result, the nuclear 
virial term in (\ref{e:PeVir}) disappears in the pressure. 
On the other hand, the second equation (\ref{e:Pdev2}) leads to a pressure formula in the form:
\begin{eqnarray}
3P_eV&=&-3V\left.\frac{\partial\tilde F}{\partial V}\right|_{Tn({\bf r})}
=-3V\left.\frac{\partial\tilde E}{\partial V}\right|_{Sn({\bf r})}\label{e:Pe2}\\
&=&\oint_{\partial V}{\bf r}\cdot({\bf\textsf P}^0_{\rm K}+
{\bf\textsf P}_{\rm xc})\cdot d{\bf S}\,. \label{e:PeVir2}
\end{eqnarray}
This relation is derived below by noting that the intrinsic 
internal energy is written 
as $\tilde E=T_s+\tilde E_{\rm es}+E_{\rm xc}$. 
At a first step, we get the exchange-correlation pressure in the following way (see Appendix):
\begin{eqnarray}
3V\left.\frac{\partial E_{\rm xc}}{\partial V}\right|_{SN\{{\bf R}_\alpha\}}
\!\!&=&-\!\int {\rm tr} {\bf\textsf P}_{\rm xc}d{\bf r}\\
\!\!\!\!&=&3V\left.\frac{\partial E_{\rm xc}}{\partial V}\right|_{Sn({\bf r})}
\!\!+3V\int \left.\frac{\delta E_{\rm xc}}{\delta n({\bf r})}\right|_{SV}
\left.\frac{dn({\bf r})}{dV}\right|_{SN\{{\bf R}_\alpha\}}\!\!\!\!d{\bf r}\\
&=&3V\left.\frac{\partial E_{\rm xc}}{\partial V}\right|_{Sn({\bf r})}
+\int_V n({\bf r}){\bf r}\cdot\nabla \left.\frac{\delta E_{\rm xc}}
{\delta n({\bf r})}\right|_{SV}d{\bf r}\,, \label{e:3VExc}
\end{eqnarray}
with aids of (\ref{e:fdndV}) and (\ref{e:dExcdlam}). 
Hence, on use of (\ref{e:3VExc}) and (\ref{e:PxcDef}), 
the exchange-correlation pressure is derived as 
\begin{eqnarray}
-3V\left.\frac{\partial E_{\rm xc}}{\partial V}\right|_{Sn({\bf r})}
=\int_V {\rm tr} {\bf\textsf P}_{\rm xc}d{\bf r}
+\int_V {\bf r}\cdot\nabla\cdot{\bf\textsf P}_{\rm xc} d{\bf r}
=\oint_{\partial V}{\bf r}\cdot{\bf\textsf P}_{\rm xc}\cdot d{\bf S}\,. 
\label{e:3vExc3}
\end{eqnarray}
Since $d{\tilde E_{\rm es}}/{dV}|_{Sn({\bf r})}=0$ and 
the kinetic pressure due to $T_s$ is given by (\ref{e:p0V3}), 
the electron pressure is expressed finally as (\ref{e:PeVir2}) 
with combined use of (\ref{e:3vExc3}).

In the equilibrium, the occupation probability $f_i$ fulfills the following 
 conditions
\begin{equation}\label{e:dOMdfi0}
\left.\frac{\partial\Omega}{\partial f_i}\right|_{TV\mu U({\bf r})}
=\left.\frac{\partial F}{\partial f_i}\right|_{TVNU({\bf r})}
=\left.\frac{\partial E}{\partial f_i}\right|_{SVNU({\bf r})}
=\left.\frac{\partial\Omega_s}{\partial f_i}\right|_{TV\mu_0 U_{\rm eff}({\bf r})}=0\,.
\end{equation}
This can be verified below. In the expression of thermodynamic potential, $\Omega=T_s-TS_s+(E_{\rm es}+{\cal F}_{\rm xc})-\mu N$, the derivative of the first term is written as
\begin{equation}\label{e:dTsdfi}
\frac{dT_s}{d f_i}=\left.\frac{\partial\Omega_s}{\partial f_i}\right|_{T\mu_0 U_{\rm eff}}+T\frac{dS_s}{d f_i}+
\int \gamma_{\rm eff}({\bf r})\frac{dn({\bf r})}{d f_i}d{\bf r}\,,
\end{equation}
with use of (\ref{e:dTsdlam}), and the third term yields 
\begin{eqnarray}
\left.\frac{d}{df_i}[E_{\rm es}+{\cal F}_{\rm xc}]\right|_{TVU({\bf r})}
&=&\int \left.\frac{\delta[E_{\rm es}+{\cal F}_{\rm xc}]}{\delta n({\bf r})}
\right|_{TVU({\bf r})}\frac{dn({\bf r})}{df_i}d{\bf r}\label{e:dEesxcDfi}\\
&=&\int [U_{\rm eff}({\bf r})+\mu_{\rm I}]\frac{dn({\bf r})}{df_i}d{\bf r}\,,
\end{eqnarray}
because of $d[\tilde E_{\rm es}\!+\!{\cal F}_{\rm xc}]/df_i|_{n({\bf r})}\!=\!0$.
Therefore, the derivative of $\Omega$ with respect to $f_i$ becomes
\begin{eqnarray}
\left.\frac{\partial\Omega}{\partial f_i}\right|_{TV\mu U({\bf r})}
&=&\left.\frac{\partial\Omega_s}{\partial f_i}\right|_{TV\mu_0 U_{\rm eff}({\bf r})}
+\int \gamma_{\rm eff}({\bf r})\frac{dn({\bf r})}{d f_i}d{\bf r}\\
&+&\int [U_{\rm eff}({\bf r})+\mu_{\rm I}]\frac{dn({\bf r})}{df_i}d{\bf r}
-\mu \frac{dN}{df_i}=\left.\frac{\partial\Omega_s}{\partial f_i}\right|_{TV\mu_0 U_{\rm eff}({\bf r})}\,, \label{e:PartOMsfi}
\end{eqnarray}
which is equal to zero owing to (\ref{e:dOMdfi}).
Here, it is important to notice in this proof that the constant term in the 
effective potential $U_{\rm eff}({\bf r})$ should be $-\mu_{\rm I}$ to fulfill (\ref{e:dOMdfi0}). 
Similarly, other relations in (\ref{e:dOMdfi0}) 
can be proved with help of (\ref{e:dTsdlam}). 
In this way, 
we can prove that the equilibrium conditions for the density distribution function $n({\bf r})$ and the Kohn-Sham eigenvalues $\epsilon_i$ to fulfill are 
given, respectively, by 
\begin{equation}\label{e:dOsn} 
\left.\frac{\delta\Omega}{\delta n({\bf r})}\right|_{TV\mu U({\bf r})}
\!\!\!=\left.\frac{\delta F}{\delta n({\bf r})}\right|_{TVNU({\bf r})}
\!\!\!=\left.\frac{\delta E}{\delta n({\bf r})}\right|_{SVNU({\bf r})}
\!\!\!=\left.\frac{\delta\Omega_s}{\delta n({\bf r})}\right|_{TV\mu_0 U_{\rm eff}({\bf r})}\!\!\!=0\,,
\end{equation}
and 
\begin{equation}\label{e:dOsep}
\left.\frac{\partial\Omega}{\partial \epsilon_i}\right|_{T\mu U({\bf r})}
\!\!=\left.\frac{\partial F}{\partial \epsilon_i}\right|_{TNU({\bf r})}
\!\!=\left.\frac{\partial E}{\partial \epsilon_i}\right|_{SNU({\bf r})}
\!\!=\left.\frac{\partial\Omega_s}{\partial \epsilon_i}\right|_{T\mu_0U_{\rm eff}({\bf r})}
\!\!=f_i.
\end{equation}

Finally, we derive a finite-temperature generalization of the Janak theorem~\cite{Janak78} in the form:
\begin{equation}\label{e:JanakT}
\left.\frac{dE}{df_i}\right|_{VU({\bf r})}=\epsilon_i+\mu_{\rm I}\,.
\end{equation}
which is different from original Janak's theorem in that the ground-state energy is 
replaced by the internal energy $E$ in Eq.~(\ref{e:JanakT}). 
This can be easily proved with help of Eq.~(\ref{e:dTsdlam}), 
which is reduced to 
\begin{equation}\label{e:dTsdfiV}
\left.\frac{dT_s}{d f_i}\right|_V=
T\frac{dS_s}{d f_i}+
\int \gamma_{\rm eff}({\bf r})\frac{dn({\bf r})}{d f_i}d{\bf r}\,,
\end{equation}
because of ${\partial\Omega_s}/{\partial f_i}|_{TV\mu_0 U_{\rm eff}}=0$.
Since $E\!=\!T_s\!+\!E_{\rm es}\!+\!E_{\rm xc}$, there results 
\begin{eqnarray}
\left.\frac{dE}{df_i}\right|_{VU({\bf r})}
\!\!\!\!&=&T\frac{dS_s}{d f_i}+
\int \gamma_{\rm eff}({\bf r})\frac{dn({\bf r})}{d f_i}d{\bf r}
+\int\left.\frac{\delta[E_{\rm es}+E_{\rm xc}]}{\delta n({\bf r})}
\right|_{SVU({\bf r})}\!\!\!\frac{dn({\bf r})}{df_i}d{\bf r}\label{e:dEdfiV}\\
&=&-\frac1\beta \ln \frac{f_i}{1-f_i}+\mu=\epsilon_i+\mu_{\rm I}\,.
\end{eqnarray}

\section{Summary and discussion}
In the first place, we described thermodynamics in the DF theory 
for a neutral electron-nucleus mixture in \S~\ref{s:DF} 
in a systematic manner. Next, 
we derived two fundamental relations, (\ref{e:dTsdlam}) 
and (\ref{e:dTsdlamUeff}), in the non-interacting system, which lead to the 
thermodynamical relations in the interacting system described in terms of quantities of the non-interacting system and the exchange-correlation effect. 
Thus, with combined of use of (\ref{e:dnUeff}), the relation (\ref{e:dTsdlam}) 
generates the pressure formula (\ref{e:forceT1}) as a force theorem, 
the familiar Fermi distribution law~\cite{Callaway84} (\ref{e:dOMdfi0}), a finite-temperature generalization of 
Janak's theorem (\ref{e:JanakT}) and some other equilibrium conditions, 
(\ref{e:dOsn}) and (\ref{e:dOsep}), which are translated into 
those of the non-interacting system, respectively.
On the other hand, the relation (\ref{e:dTsdlamUeff}) provides two 
types of force expressions, (\ref{e:forceT2}) and (\ref{e:AFT2}) 
(Andersen's force theorems,~\cite{Andersen85,MackAnd85,Heine80,Skriver84}), in conjunction with another type of 
force theorem (\ref{e:forceDE}), which is equivalent to (\ref{e:forceT2}).
It is interesting to note that thermodynamic 
relations, (\ref{e:Pdev1}) and (\ref{e:Pdev2}), provide
two different but equivalent expressions for the pressure: 
the one, (\ref{e:PeVir}) in the volume-integral form and
the other, (\ref{e:PeVir2}) in the surface-integral form, respectively.

In the standard DF theory,\cite{KohnSham65,Rajago80,KohnVashi83,Callaway84} the effective external potential is defined as
\begin{equation} 
\bar U_{\rm eff}({\bf r})=\left.\frac{\delta [E_{\rm es}+{\cal F}_{\rm xc}]}{\delta n({\bf r})}\right|_{TVU({\bf r})}
=\left.\frac{\delta [E_{\rm es}+{E}_{\rm xc}]}{\delta n({\bf r})}\right|_{SVU({\bf r})}\,,
\end{equation}
instead of the definition (\ref{e:Ueffdn}), where the interaction part of the 
chemical potential $\mu_{\rm I}$ is subtracted to give the effective external 
potential. Since the eigenvalues 
$\bar \epsilon_i$ for $\bar U_{\rm eff}({\bf r})$ are related with 
those of $U_{\rm eff}({\bf r})$ as 
$\bar \epsilon_i=\epsilon_i +\mu_{\rm I}$, the formulas for 
$\bar U_{\rm eff}({\bf r})$ are obtained by replacing 
as $\epsilon_i\!-\!\mu_0=\bar\epsilon_i\!-\!\mu$, and $\mu_0\!-\!U_{\rm eff}({\bf r})=\mu\!-\!\bar U_{\rm eff}({\bf r})$ from our results.
For example, the occupation probability for a state $\bar\epsilon_i$ is 
given by $f_i=f(\bar\epsilon_i\!-\!\mu)=f(\epsilon_i\!-\!\mu_0)$ with use of 
the chemical potential $\mu$ of the real system instead of $\mu_0$.
Also, the Janak theorem is written in the usual form: ${dE}/{df_i}|_V=\bar\epsilon_i$. 
On the other hand, the relation~\cite{KohnVashi83}, 
$\bar\epsilon_{\rm max}=\mu$, for the occupied maximum level $\bar\epsilon_{\rm max}$ at $T=0$ is expressed as $\epsilon_{\rm max}=\mu_0$ for $U_{\rm eff}({\bf r})$.

When we obtain the free-energy $F$ of the electrons of a neutral electron-nucleus mixture on the basis of the DF theory with aids of the adiabatic approximation,
the total pressure of this system is written as~\cite{Chihara01}
\begin{eqnarray}
3PV&=&3(P_e+P_{\rm n})V \nonumber \\
&=&\left\langle-3V\left.\frac{\partial F}{\partial V}
\right|_{TN\{{\bf R}_\alpha\}}\right\rangle_{\rm n}+2\langle \hat T_{\rm n}\rangle_{\rm n}
+\sum_\alpha\langle{\bf R}_\alpha\cdot{\bf F}_\alpha\rangle_{\rm n}\\
&=&2\langle \hat T_{\rm n}\rangle_{\rm n}+2\langle T_s\rangle_{\rm n}
+\langle{E}_{\rm es}\rangle_{\rm n}
+\int_V \langle{\rm tr} {\bf\textsf P}_{\rm xc}\rangle_{\rm n} d{\bf r}
\end{eqnarray}
with the force on $\alpha$-nucleus: 
${\bf F}_\alpha=-\nabla_\alpha F|_{TVN}=-\nabla_\alpha E_{\rm es}$.
Here, the bracket with a suffix n denotes the average concerning the nuclear variables defined as
\begin{equation}
\langle A \rangle_{\rm n}\equiv \frac{{\rm Tr}_{\rm n}\{
\exp[-\beta(\hat T_{\rm n}+F)]A\}}{{\rm Tr}_{\rm n}\{
\exp[-\beta(\hat T_{\rm n}+F)]\}}\,,
\end{equation}
where $\rm Tr_{\rm n}$ means a trace using a complete set of eigenfunctions for 
the nuclear Hamiltonian, $\hat T_{\rm n}+F\{\hat{\bf R}_\alpha\}$, 
with the nuclear kinetic operator $\hat T_{\rm n}$. 
In calculating the pressure for a solid, 
the nuclear pressure $P_{\rm n}$ is usually neglected.

\appendix
\section{} 
When $F$ is a function of the independent variables $\{x_1, x_2,\ldots, x_n\}$,
the derivative of $F$ with respect to some parameter $\lambda$ is expressed as
\begin{equation}
\frac{dF}{d\lambda}=\sum_{i=1,n}\left.\frac{\partial F}{\partial x_i}
\right|_{{\bar x}_i}\frac{dx_i}{d\lambda}
\end{equation}
with $\bar x_i$ denoting a set of variables $\{x_1,\ldots,x_n\}$ except $x_i$. Thus, there results 
\begin{eqnarray}\label{e:diffParm}
\left.\frac{dF}{d\lambda}\right|_{x_1\ldots x_{m-1}}=\left.\frac{dF}{d\lambda}\right|_{x_1\ldots x_m}+\left.\frac{\partial F}{\partial x_m}\right|_{{\bar x}_m}
\frac{dx_m}{d\lambda}\,,
\end{eqnarray}
because of ${dF}/{d\lambda}|_{x_1\ldots x_m}=\sum_{i=m+1,n}{\partial F}/{\partial x_i}|_{{\bar x}_i}\cdot{dx_i}/{d\lambda}$.

On the other hand, when we transform the independent variables $\{x_1, x_2,\ldots, x_n\}$ to $\{x_1, x_2,\ldots, y\}$, we get a formula akin to 
but different from (\ref{e:diffParm}):
\begin{equation}
\left.\frac{\partial F}{\partial x_1}\right|_{{\bar x}_1}
\!\!=\left.\frac{\partial(F,x_2,\ldots,x_n)}{\partial (x_1,x_2,\ldots,y)}
\right/\frac{\partial(x_1,x_2,\ldots,y)}{\partial (x_1,x_2,\ldots,x_n)}
=\left.\frac{\partial F}{\partial x_1}\right|_{x_2\ldots y}\!\!\!\!\!+\left.\frac{\partial F}{\partial y}\right|_{\bar x_n}
\!\!\left.\frac{\partial y}{\partial x_1}\right|_{{\bar x}_1}.
\end{equation}

\end{document}